\documentclass{amsart}
\usepackage{graphicx} 
\usepackage{longtable}
\usepackage{graphicx}
\usepackage{subcaption}
\usepackage{mathtools}
\usepackage{color}

\usepackage{cite}
\usepackage{url}
\usepackage{amsmath}
\usepackage{amssymb}
\usepackage{amsthm}
\usepackage{setspace}
\usepackage[english]{babel}
\usepackage{mathrsfs}
\usepackage{mathtools}
\usepackage[utf8]{inputenc}
\usepackage{xcolor}
\usepackage{mathrsfs}
\usepackage{enumerate}

\newtheorem{definition}{Definition}[subsection]
\newtheorem{remark}{Remark}[subsection]
\newtheorem{theorem}{Theorem}[subsection]

\newtheorem{example}{Example}[subsection]

\renewcommand{\epsilon}{\varepsilon}
\renewcommand{\Im}{\mathrm{Im}}

\newcommand{\dd}{\,\mathrm{d}}  
\newcommand{\ee}{\mathrm{e}} 

\newcommand{\sgn}{\mathrm{sgn}}
\DeclareMathOperator{\Dom}{Dom}
\DeclareMathOperator{\Ran}{Ran}
\DeclareMathOperator{\Ker}{Ker}

\DeclareMathOperator{\supp}{supp}

\newcommand{\N}{\mathbb{N}}
\newcommand{\C}{\mathbb{C}}
\newcommand{\R}{\mathbb{R}}

\title{Non-local relativistic $\delta$-shell interactions}
\author{Luk\'{a}\v{s} Heriban}
\address{Department of Mathematics\\ Faculty of Nuclear Sciences and Physical Engineering \\ Czech Technical University in Prague\\ Trojanova 13\\ 120 00 Prague\\ Czechia}
\email{heribluk@fjfi.cvut.cz}
\author{Mat\v{e}j Tu\v{s}ek}
\address{Department of Mathematics\\ Faculty of Nuclear Sciences and Physical Engineering \\ Czech Technical University in Prague\\ Trojanova 13\\ 120 00 Prague\\ Czechia}
\email{matej.tusek@fjfi.cvut.cz}
\date{November 5, 2023}

\begin{document}

\begin{abstract}
In this paper, new self-adjoint realizations of the Dirac operator in dimension two and three are introduced. It is shown that they may be associated with the formal expression $\mathcal{D}_0+|F\delta_\Sigma\rangle\langle G\delta_\Sigma|$, where $\mathcal{D}_0$ is the free Dirac operator, $F$ and $G$ are matrix valued coefficients, and $\delta_\Sigma$ stands for the single layer distribution supported on a hypersurface $\Sigma$, and that they can be understood as limits of the Dirac operators with scaled non-local potentials. Furthermore, their spectral properties are analysed.
\end{abstract}

\maketitle 

\section{Introduction} 
Let $\Omega$ be an open bounded simply connected subset of $\R^n$, $n=2,\,3$, with the  Lipschitz smooth boundary $\Sigma$. In the present paper, we will deal with the $n$-dimensional Dirac operator perturbed by the formal \emph{non-local $\delta$-shell} potential
\begin{equation} \label{eq:formal_FG}
|F\delta_\Sigma\rangle\langle G\delta_\Sigma|,
\end{equation}
where $F,G\in L^2(\Sigma;\C^{N\times N})$ are matrix-valued functions with $N:=2^{\lceil \frac{n}{2} \rceil}$ and $\delta_\Sigma$ stands for the single layer distribution supported on $\Sigma$. Identifying the \emph{bra-vector} with the adjoint of the distribution action and extending naturally the usual operations with distributions, we put, for $\varphi\in\mathscr{D}(\R^n;\C^N)$,
\begin{equation} \label{eq:FGaction}
|F\delta_\Sigma\rangle\langle G\delta_\Sigma|\varphi:=F(\delta_\Sigma,G^*\varphi)\delta_\Sigma=F\int_\Sigma (G^*\varphi)\,\delta_\Sigma\in\mathscr{D}'(\R^n;\C^N),
\end{equation}
i.e.,
\begin{equation*}
(|F\delta_\Sigma\rangle\langle G\delta_\Sigma|\varphi,\tilde\varphi)=\int_\Sigma\sum_{i=1}^N \Big(F\int_\Sigma (G^*\varphi)\Big)_i\tilde\varphi_i\in\C\quad (\forall\tilde\varphi\in\mathscr{D}(\R^n;\C^N)).
\end{equation*}
Note that choosing $F=I,\, G=A^*=\text{const.}$, the expression \eqref{eq:formal_FG} is reduced to
\begin{equation} \label{eq:formal_A}
A|\delta_\Sigma\rangle\langle\delta_\Sigma|,
\end{equation}
which acts as
\begin{equation} \label{eq:formal_A_def}
A|\delta_\Sigma\rangle\langle\delta_\Sigma|\varphi:=A(\delta_\Sigma,\varphi)\delta_\Sigma\in\mathscr{D}'(\R^n;\C^N).
\end{equation}

We will show that the Dirac operator with the potential \eqref{eq:formal_FG} is described by a non-local transmission condition along $\Sigma$ for the functions in the operator domain, which relates the difference of the inner and the outer trace with respect to $\Omega$ and its complement, respectively, to the integral over $\Sigma$ of the their average, see \eqref{eq:TC}. To our best knowledge, such an operator has not been considered before. However, it is quite common to use \eqref{eq:formal_A} with $A=A^*$ as a formal expression for the point interaction in the one-dimensional setting (and, in the non-relativistic setting, also for the point interaction in dimensions two and three). Recall that then
\begin{equation*}
|\delta\rangle\langle\delta|\varphi=(\delta,\varphi)\delta=\varphi(0)\delta=\varphi\delta=:\delta \varphi,
\end{equation*}
i.e., the "projection on the Dirac $\delta$-distribution" yields exactly the same result as the "multiplication by the Dirac $\delta$-distribution". This is not true for the single layer distribution, because
\begin{equation*}
A\delta_\Sigma\varphi:=A\varphi\delta_\Sigma\in\mathscr{D}'(\R^n;\C^N)
\end{equation*}
is clearly different from \eqref{eq:formal_A_def}.

The Dirac operator with the formal potential $A\delta_\Sigma$ is described by a local transmission condition along $\Sigma$ and has been studied intensively during recent years, see, e.g., \cite{ArMaVe_14,BeExHoLo_19,Beh Hol Our Pan,Be_22}. In particular, an important question of regular approximations was addressed in \cite{Mas Piz, Cas Lot Mas Tus,BeHoSt_23}. It was observed that given a scaled regular potential $V_\varepsilon$ that converges to $\delta_\Sigma$ as $\varepsilon\to 0+$, the Dirac operator with the potential $A V_\varepsilon$ converges to the Dirac operator with the formal potential $\tilde A \delta_\Sigma$, where, except for special cases,  $\tilde A\neq A$, i.e., the coupling constants have to be renormalized. The same surprising effect occurs in the one-dimensional setting, too \cite{Seb, Tus}. Nevertheless, when non-local approximations of the type $|V_\varepsilon\rangle\langle V_\varepsilon|$ are used, no renormalization is needed \cite{Seb, Her Tus}. When we tried to show a similar result in a higher dimensional setting, we discovered that the limit operator for the Dirac operator with the potential $A |V_\varepsilon\rangle\langle V_\varepsilon|$ cannot be associated with the formal potential $\tilde A\delta_\Sigma$ for any $\tilde A$. Instead, we found out that it corresponds to the Dirac operator with the formal potential $A|\delta_\Sigma\rangle\langle\delta_\Sigma|$.

The paper is organized as follows. In Section \ref{sec:prelim}, we will introduce notations, present basic facts about the free Dirac operator and tubular neighbourhoods of $\Sigma$, and  summarize very briefly the theory of generalized boundary triples, that constitutes the cornerstone of our analysis. More concretely, we will rely on the generalized boundary triple  for the Dirac operator that has been developed very recently in \cite{Beh Hol Ste Ste}. After recalling this triple in Section \ref{sec:delta_shell}, the Dirac operator with the potential \eqref{eq:formal_FG} will be introduced rigorously, the condition on its self-adjointness will be derived, and its spectrum will be investigated. 
In Section \ref{sec:approx}, approximations of the non-local $\delta$-shell potentials will be constructed by means of scaled non-local finite-rank potentials. We will prove that the resolvents of the respective Dirac operators converge uniformly, which implies the convergence of spectra and eigenfunctions.

\section{Preliminaries} \label{sec:prelim}

\subsection{Notations}
By $L^2(\mathcal{M};\mathcal{G})$ we denote the space of functions (after the usual factorization) with values in the Banach space $\mathcal{G}$ defined on $\mathcal{M}$ for which the $2$nd power of the norm on $\mathcal{G}$ is integrable, where for $\mathcal{M}$ being an open subset of $\R^n$ or a hypersurface embedded in $\R^n$, we integrate with respect to the $n$-dimensional Lebesgue measure or the surface measure induced by the embedding, respectively.   Similarly, for $s\in\mathbb{R}$ we denote by $H^s(\mathcal{M};\C^N)$  the space of $\C^N$-valued functions on $\mathcal{M}$ such that each of their components belongs to the standard $L^2$-based Sobolev space $H^s(\mathcal{M})$. We use the symbol $\langle\cdot,\cdot\rangle$  for the dot product (conjugate linear in the first argument) on $L^2(\R^n;\C^N)$ and also as a natural abbreviation for the following integrals
\begin{align*}
\langle E,\psi\rangle &:= \int_{\R^n} E^*(x)\psi(x)\dd x\in\C^N,\\ 
\langle E,H\rangle &:= \int_{\R^n} E^\ast(x)H(x)\dd x\in\C^{N\times N},
\end{align*}
where $\psi\in L^2(\R^n;\C^N)$ and $E,\, H\in L^2(\R^n;\C^{N\times N})$. Then by $|E\rangle\langle H|$ we understand the following finite rank operator in $L^2(\R^n;\C^N)$,
\begin{equation*}
|E\rangle\langle H|\psi:=E\langle H,\psi\rangle=E\int_{\R^n}H^*(x)\psi(x)\dd x.
\end{equation*}

The action of a distribution $f$ on a test function $\varphi$ is denoted by $(f,\varphi)$. This bracket is linear in both arguments. If $K$ is an integral operator then we  write $K(x,y)$ for its kernel. The surface measure on $\Sigma$ is denoted by $\dd\sigma$. We always adopt the convention $(\forall w\in\mathbb{C}\setminus [0,+\infty))(\Im \sqrt{w}>0).$
 We use the standard definition of the Pauli matrices,
\begin{equation*}
\sigma_1 = \begin{pmatrix} 
0 & 1\\
1 & 0
\end{pmatrix},\,
\sigma_2 = \begin{pmatrix} 
0 & -i\\
i & 0
\end{pmatrix},\,
\sigma_3 = \begin{pmatrix} 
1 & 0\\
0 & -1
\end{pmatrix},
\end{equation*}
and by $I_N$ we denote the $N\times N$ identity matrix. 

\subsection{Free Dirac operator}
Let $m\in\R$, $n\in\{2,3\}$, and  $N=2^{\lceil \frac{n}{2} \rceil}$. For $n=2$, we put
\begin{equation*}
\alpha_1=\sigma_1,\, \alpha_2=\sigma_2,\, \alpha_0=\sigma_3,
\end{equation*}
whereas for $n=3$, we put
\begin{equation*}
\alpha_k = \begin{pmatrix}
    0 & \sigma_k\\
    \sigma_k & 0
\end{pmatrix}\, (\forall k \in \{1,2,3\}),\,\, \alpha_0 = \begin{pmatrix}
    I_2 & 0\\
    0 & -I_2
\end{pmatrix}.
\end{equation*}
Finally, let $\mathcal{D}_0$ be the differential expression that acts on $\C^N$-valued functions of $n$-variables as 
\begin{equation*}
\mathcal{D}_0:=-i(\alpha\cdot\nabla)+m\alpha_0=-i\sum_{k=1}^n\alpha_k\frac{\partial}{\partial x_k}+m\alpha_0.
\end{equation*}
Then the $n$-dimensional free Dirac operator with the mass term $m$ is the following  operator in $L^2(\R^n;\C^N)$,
\begin{equation*}
\begin{split}
     \Dom (D_0) =& H^1 (\mathbb{R}^n; \mathbb{C}^N),\\
     D_0 \psi =& \mathcal{D}_0 \psi.
\end{split}
\end{equation*}

It is well known that the operator $D_0$ is self-adjoint and its spectrum is purely absolutely continuous and consists of 
$$\sigma(D_0) = \sigma_{\text{ac}}(D_0)  = (-\infty,-|m|]\cup[|m|,+\infty),$$ 
cf. \cite{Thaller}. For $z\in\C\setminus\sigma(D_0)$, the integral kernel of the resolvent $(D_0-z)^{-1}$ may be computed from the integral kernel of the resolvent of the Laplacian, employing the relation 
$$(D_0-z)(D_0+z)=I_N(-\Delta+m^2-z^2).$$ 
Explicitly, it is given by $(D_0-z)^{-1}(x,y)=:R_z(x-y)$ with
\begin{equation*}\label{resolvent 2D}
    R_z(x)= \frac{k(z)}{2\pi}K_1(-ik(z)|x|)\frac{(\alpha\cdot x)}{|x|}+\frac{1}{2\pi}K_0(-ik(z)|x|)(zI_2 +m\alpha_0)
\end{equation*}
and
\begin{equation*}\label{resolvent 3D}
    R_z(x)=\left(zI_4+m\alpha_0+(1-ik(z)|x|)\frac{i(\alpha\cdot x)}{|x|^2}\right)\frac{1}{4\pi|x|}\ee^{ik(z)|x|}
\end{equation*}
for the dimension two and three, respectively. Here, $K_j$ stands for the modified Bessel function of the second kind and $k(z) := \sqrt{z^2-m^2}.$

\subsection{Tubular neighbourhoods of hypersurfaces} \label{sec:tubular}
If $\Omega$ is as above but now with $C^2$-smooth boundary $\Sigma$, then we may construct  tubular neighbourhoods of $\Sigma$ in a standard way, cf. \cite{KrRaTu_15}. Namely, given $\varepsilon>0$, we define the $\varepsilon$\emph{-tubular neighbourhood} $\Sigma_\varepsilon$ of $\Sigma$ as the image of the mapping
\begin{equation*}
\mathscr{L}_\varepsilon:\, \Sigma\times(-1,1)\to\R^n:\, \left\{(x_\Sigma,u)\mapsto x_\Sigma+\varepsilon u \nu(x_\Sigma)\right\},
\end{equation*}
where $\nu(x_\Sigma)$ stands for the unit normal vector pointing outwards of $\Omega$, i.e., 
\begin{equation} \label{eq:epsNeigh}
\Sigma_\varepsilon:=\mathscr{L}_\varepsilon(\Sigma\times(-1,1)). 
\end{equation}
Under our assumptions, the principal curvatures $K_\mu,\, \mu=1,\ldots, n-1,$  of $\Sigma$ are continuous functions on the compact set $\Sigma$. Therefore, $\mathscr{L}_\varepsilon$ is a local diffeomorphism for all $\varepsilon$ sufficiently small, cf. formula \eqref{eq:VolElem} below. Moreover, the definition of a $C^k$-smooth domain combined with the compactness of $\Sigma$ yield that $\mathscr{L}_\varepsilon:\Sigma\times(-1,1)\to \Sigma_\varepsilon$ is bijective  for all $\varepsilon$ below a certain threshold.

We may view $\Sigma$ as a Riemannian manifold equipped with the metric $g$ induced by the embedding into $\R^n$ and $\Sigma_\varepsilon$ as a Riemannian manifold with the metric induced by $\mathscr{L}_\varepsilon$.  Then the \emph{volume element} on $\Sigma_\varepsilon$ obeys
\begin{equation} \label{eq:VolElem}
 \dd\Omega_\varepsilon =\varepsilon\left[
  1+\sum_{\mu=1}^{n-1}(-\varepsilon u)^{\mu}\binom{d-1}{\mu}K_{\mu}
  \right]\, \dd\sigma \wedge \dd u=:\varepsilon w_\varepsilon \, \dd\sigma \wedge \dd u, 
\end{equation}
where $\dd\sigma := (\det{g^{-1}})^{1/2} \, \dd x^1 \wedge \dots \wedge \dd x^{n-1}$ with $x^{\mu}$ being, for the moment, local coordinates on $\Sigma$ is the volume element on $\Sigma$. Note that the function $w_\varepsilon=w_\varepsilon(x_\Sigma,u)$ obeys
\begin{equation} \label{eq:wAsy}
 w_\varepsilon=1+\mathcal{O}(\varepsilon)
\end{equation}
uniformly on $\Sigma\times(-1,1)$ as $\varepsilon\to 0+$.

\subsection{Generalized boundary triple}\label{section: triples}
Below, we will summarize basic definitions and results concerning the quasi and the generalized boundary triples. We will mainly follow \cite{Beh Hol Ste Ste} in our exposition; original statements with complete proofs may be found in \cite{DeMa_91,DeMa_95,BeLa_07,BeLa_12}. Throughout this section, $S$ is assumed to be a densely defined closed symmetric operator in a Hilbert space $\mathcal{H}$ and $T$ is a linear operator such that $\overline{T}=S^*$. The quasi/generalized boundary triples provide powerful tools for studying certain restrictions of $T$ (which turn out to be extensions of $S$).

Later, in our particular setting, $S$ will be a restriction of the free Dirac operator $D_0$ to a  subspace of functions that vanish along the hypersurface $\Sigma$ and certain extensions of $S$ constructed using a generalized boundary triple will be identified with the Dirac operators perturbed by the non-local $\delta$-shell interaction \eqref{eq:formal_FG}.

\begin{definition}\label{def: boundary triple}
Let $T$ be  such that $\overline{T}=S^\ast$. A triple $(\mathcal{G},\Gamma_0,\Gamma_1)$ consisting of a Hilbert space $\mathcal{G}$ and linear mappings $\Gamma_0,\Gamma_1:\Dom T \to \mathcal{G}$ is called a quasi boundary triple for $S^\ast$ if the following holds:
\begin{enumerate}[(i)]
    \item For all $f,g\in\Dom T$, $\langle Tf,g\rangle_\mathcal{H} - \langle f,Tg\rangle_\mathcal{H}= \langle \Gamma_1 f,\Gamma_0 g\rangle_\mathcal{G} - \langle \Gamma_0 f,\Gamma_1 g\rangle_\mathcal{G}.$
    \item The range of $(\Gamma_0,\Gamma_1)$ is dense in $\mathcal{G}\times\mathcal{G}$.
    \item  The restriction $T_0:=T\restriction{\Ker \Gamma_0}$ is a self-adjoint operator in $\mathcal{H}$.
\end{enumerate}
\end{definition}
If conditions (i) and (iii) hold, and the mapping $\Gamma_0:\Dom T \to \mathcal{G}$ is surjective, then $(\mathcal{G},\Gamma_0,\Gamma_1)$ is called generalized boundary triple. 
Note that \cite[Lem. 6.1]{DeMa_95} implies that every generalized boundary triple is also a quasi boundary triple.

\begin{definition}
    Let $S,\, T$ be as above,  $(\mathcal{G},\Gamma_0,\Gamma_1)$ be a quasi boundary triple for $S^\ast$, and $T_0=T\restriction{\Ker\Gamma_0}$. Then the associated $\gamma$-field and the Weyl function $M$ are defined by 
    $$\rho(T_0)\ni z \mapsto \gamma(z) = (\Gamma_0\restriction{\Ker(T-z)})^{-1}$$
    and 
    $$\rho(T_0)\ni z \mapsto M(z)= \Gamma_1(\Gamma_0\restriction{\Ker(T-z)})^{-1}.$$
\end{definition} 

For a linear operator $B$ in $\mathcal{G}$, we put
\begin{equation}\label{T_B}
T_B = T\restriction{\Ker(\Gamma_0+B\Gamma_1)}.
\end{equation}
Since $\Dom{S}=\ker\Gamma_0\cap\ker\Gamma_1$ by \cite[Prop. 2.2]{BeLa_07}, $S\subset T_B$. The following theorem yields an eigenvalue condition for $T_B$, an alternative description of $\Ran(T_B-z)$, which may be used in the proof of self-adjointness of $T_B$, and a Krein-like formula for the resolvent of $T_B$.

\begin{theorem}\label{Theo: Boundary triple}
    Let $S,\, T$ be as above, $(\mathcal{G},\Gamma_0,\Gamma_1)$ be a quasi boundary triple for $S^*$, $T_0=T\restriction{\Ker\Gamma_0}$, and  $\gamma$ and $M$ denote the associated $\gamma$-field and the Weyl function, respectively. Finally, let $T_B$ be given by \eqref{T_B}. Then the following holds for all $z\in\rho(T_0)$:
    \begin{enumerate}[(i)]
        \item $z\in\sigma_{\mathrm p} (T_B)$ if and only if $0\in\sigma_{\mathrm p}(I+BM(z))$. Moreover, $$\Ker(T_B -z) = \{\gamma(z)\psi\mid \psi\in\Ker (I+BM(z))\}.$$
        \item If $z\notin \sigma_{\mathrm p}(T_B)$, then $g\in\Ran (T_B-z)$ if and only if $B\gamma(\overline{z})^\ast g\in \Ran (I+BM(z)).$
        \item If $z\notin\sigma_{\mathrm p}(T_B)$, then 
        \begin{equation} \label{eq:Krein}
            (T_B-z)^{-1}g = (T_0-z)^{-1}g-\gamma(z)(I+BM(z))^{-1}B\gamma(\overline{z})^\ast g
        \end{equation}
        holds for all $g\in\Ran (T_B-z)$.
    \end{enumerate}
\end{theorem}

\section{Non-local relativistic delta shell interactions} \label{sec:delta_shell}
Recall that we assume $\Sigma$ to be the Lipschitz smooth boundary of an open bounded simply connected  set $\Omega\equiv\Omega_+ \subset \mathbb{R}^n,\, n=2,\,3$. Denote \emph{the outer domain} $\R^n\setminus\overline{\Omega_+}$ by $\Omega_-$. Then we can write the Euclidean space as the disjoint union $\mathbb{R}^n=\Omega_+ \cup \Sigma \cup \Omega_-$. Also recall that we denote by $\nu(x_\Sigma)$ the unit normal vector at $x_\Sigma\in\Sigma$ pointing outwards of $\Omega_+$. For $s\in[0,1]$, define the space 
$$H^s_\alpha(\Omega_\pm) :=\{ \psi_\pm \in H^s(\Omega_\pm; \mathbb{C}^N)\mid (\alpha\cdot\nabla)\psi_\pm \in L^2(\Omega_\pm; \mathbb{C}^N)\}.$$
It was shown in \cite[Lem. 4.1 and Cor. 4.6]{Beh Hol Ste Ste} that $\psi\in H^s_\alpha(\Omega_\pm)$ admits Dirichlet traces $\mathcal{T}_\pm$ in $H^{s-\frac{1}{2}}(\Sigma;\mathbb{C}^N).$ In particular, $\mathcal{T}_\pm\psi_\pm\in L^2(\Sigma;\C^N)$ for $\psi_\pm\in H^{\frac{1}{2}}_\alpha(\Omega_\pm)$.

Now, it is straightforward to check that the restriction $S:=D_0\restriction H_0^1(\R^n\setminus\Sigma;\C^N)$ is closed, symmetric, and densely defined. Moreover,
it follows that $S^*$ is given by
\begin{align*}
\Dom(S^*) &= \{\psi_-\oplus\psi_+\mid\, \psi_\pm \in H^0_\alpha(\Omega_\pm)\},\\
S^*(\psi_-\oplus\psi_+)&=\mathcal{D}_0\psi_-\oplus\mathcal{D}_0\psi_+,
\end{align*}
cf. \cite[Prop. 3.1]{BeHo_20}. It was proved in \cite{Beh Hol Ste Ste} that $H^\frac{1}{2}_\alpha(\Omega_-)\oplus H^\frac{1}{2}_\alpha(\Omega_+)$ is an operator core of $S^*$ and that, with the choice
\begin{equation} \label{eq:T}
T:=S^*\restriction H^\frac{1}{2}_\alpha(\Omega_-)\oplus H^\frac{1}{2}_\alpha(\Omega_+),
\end{equation}
the triple $(\mathcal{G},\Gamma_0,\Gamma_1)$, where $\mathcal{G}=L^2(\Sigma;\C^N)$ and
\begin{equation} \label{Gammas}
    \begin{split} 
        \Gamma_0 \psi &= i(\alpha\cdot \nu)(\mathcal{T}_+\psi_+-\mathcal{T}_-\psi_-) :\Dom T \to L^2(\Sigma;\mathbb{C}^N),\\
        \Gamma_1 \psi &= \frac{1}{2}(\mathcal{T}_+\psi_+ +\mathcal{T}_-\psi_-):\Dom T \to L^2(\Sigma;\mathbb{C}^N),
    \end{split}    
\end{equation}
is a generalized boundary triple for $S^*$. Here, $\alpha\cdot \nu :=\sum_{k=1}^n \nu_k\alpha_k$. In the same paper, it was shown that, for $z\in\rho(D_0)$, the associated $\gamma$-field and the Weyl function are given by
\begin{equation*}
    \gamma(z)\psi(x) = \int_\Sigma R_{z}(x-y_\Sigma)\psi(y_\Sigma)\dd \sigma(y_\Sigma)\quad (\forall x\in\mathbb{R}^n\setminus \Sigma)
\end{equation*}
and
\begin{equation*}
    M(z)\psi(x_\Sigma) = \lim_{\rho\to 0+}\int_{\Sigma\setminus B(x_\Sigma,\rho)}R_{z}(x_\Sigma-y_\Sigma)\psi(y_\Sigma)\dd\sigma(y_\Sigma) \quad (\forall x_\Sigma\in\Sigma),
\end{equation*}
respectively. Moreover, $\gamma(z)$ is a bounded and everywhere defined operator from $L^2(\Sigma;\C^N)$ to $L^2(\R^n;\C^N)$ with a compact adjoint and $M(z)$ is a bounded and everywhere defined operator in $L^2(\Sigma;\C^N)$.

Now, our aim will be to identify the Dirac operator perturbed by the formal potential \eqref{eq:formal_FG} with a certain operator of the form \eqref{T_B}, where $T$ is given by \eqref{eq:T} and $\Gamma_0,\,\Gamma_1$ by \eqref{Gammas}. First, using integration by parts, one gets 
\begin{equation*}
\mathcal{D}_0(\psi_-\oplus\psi_+)=T(\psi_-\oplus\psi_+)+i(\alpha\cdot \nu)(\mathcal{T}_+ \psi_+ -\mathcal{T}_- \psi_-)\delta_\Sigma \quad (\forall \psi\equiv\psi_-\oplus\psi_+\in\Dom(T)).
\end{equation*} 
Next,  note that the right-hand side of \eqref{eq:FGaction} makes sense also for $\varphi\in L^2(\Sigma;\C^N)$. Therefore, it is natural to extend  the action of \eqref{eq:formal_FG} on $\psi\in\Dom(T)$ as follows
\begin{equation*}
|F\delta_\Sigma\rangle\langle G\delta_\Sigma|\psi:=F\int_\Sigma\Big(G^*\frac{1}{2}(\mathcal{T}_+ \psi_+ +\mathcal{T}_- \psi_-)\Big)\,\delta_\Sigma.
\end{equation*}
We see that, for $\psi\in\Dom(T)$, $(\mathcal{D}_0+|F\delta_\Sigma\rangle\langle G\delta_\Sigma|)\psi$ belongs to $L^2(\R^n;\C^N)$ if and only if 
\begin{equation} \label{eq:TC}
i(\alpha\cdot \nu)(\mathcal{T}_+ \psi_+ -\mathcal{T}_- \psi_-)+F\int_\Sigma \Big(G^* \frac{1}{2}(\mathcal{T}_+ \psi_+ +\mathcal{T}_- \psi_-)\Big)=0
\end{equation}
as an element of $L^2(\Sigma;\C^N)$. This leads us to the following definition.
\begin{definition}
By the Dirac operator with non-local $\delta$-shell interaction of the type $|F\delta_\Sigma\rangle\langle G\delta_\Sigma|$ we mean the linear operator $D_{F,G}$ in $L^2(\R^n;\C^N)$ given by 
\begin{equation*}
\begin{split}
\Dom{D_{F,G}}&=\{\psi_-\oplus\psi_+\in H^{\frac{1}{2}}_\alpha(\Omega_-)\oplus H^{\frac{1}{2}}_\alpha(\Omega_+)\mid\, \psi \text{ satisfies }\eqref{eq:TC}\},\\
D_{F,G}(\psi_-\oplus\psi_+)&=\mathcal{D}_0\psi_-\oplus\mathcal{D}_0\psi_+.
\end{split}
\end{equation*}
\end{definition}

The transmission condition \eqref{eq:TC} may be rewritten as 
\begin{equation} \label{eq:TC_B}
\Gamma_0\psi+B\Gamma_1\psi=0\quad\text{with}\quad B=|F\rangle_\Sigma\langle G|_\Sigma,
\end{equation}
where $|F\rangle_\Sigma\langle G|_\Sigma$ defined by 
$$\varphi\mapsto F\int_\Sigma(G^*\varphi)$$
is a finite rank operator in $L^2(\Sigma;\C^N)$. Note that
\begin{equation} \label{eq:FG_adjoint}
(|F\rangle_\Sigma\langle G|_\Sigma)^*=|G\rangle_\Sigma\langle F|_\Sigma.
\end{equation}
With  $B$ given in \eqref{eq:TC_B}, $D_{F,G}=T_B$. In particular, $D_{0,0}=T_0=D_0$ is the free Dirac operator. Hence, we may use Theorem \ref{Theo: Boundary triple} to study $D_{F,G}$ efficiently. We start with a result on self-adjointness.

\begin{theorem} \label{Theo:sa}
Let $F,\,G\in L^2(\Sigma;\C^{N\times N})$ be such that 
\begin{equation} \label{eq:FG=GF}
|F\rangle_\Sigma\langle G|_\Sigma=|G\rangle_\Sigma\langle F|_\Sigma.
\end{equation}
Then $D_{F,G}$ is a self-adjoint  operator.
\end{theorem}
\begin{proof}
In view of \eqref{eq:FG_adjoint}, the condition \eqref{eq:FG=GF} is equivalent to the hermiticity of $B=|F\rangle_\Sigma\langle G|_\Sigma$. The property (i) from Definition \ref{def: boundary triple} together with \eqref{eq:TC_B} then imply that the operator $D_{F,G}$ is symmetric. Hence, to prove the self-adjointness of $D_{F,G}$ it is sufficent to show that $\forall z\in\mathbb{C}\setminus\mathbb{R},\,\Ran (D_{F,G} - z) = L^2(\mathbb{R}^n;\mathbb{C}^N)$. By the symmetry of $D_{F,G}$, we also have $\sigma_{\mathrm p}(D_{F,G})\subset \mathbb{R}$. Therefore, from the point (i) of Theorem \ref{Theo: Boundary triple}, the operator $(I+B M(z))$ is injective for all $z\in\mathbb{C}\setminus\mathbb{R}$. Furthermore, $B$ is a finite rank operator and thus compact. In addition, $M(z)$ is bounded, and so we deduce that the operator $B M(z)$ is also compact in $L^2(\Sigma;\C^N)$. On top of that, $I$ is Fredholm operator with index $0$ and the same holds true for its compact perturbation $(I+B M(z))$ which implies that the operator $(I+B M(z))$ is also surjective. This yields $\Ran(D_{F,G}-z) = L^2(\mathbb{R}^n;\mathbb{C}^N)$, due to the point (ii) of Theorem \ref{Theo: Boundary triple}.
\end{proof}

\begin{remark}
The condition \eqref{eq:FG=GF} is clearly equivalent to 
\begin{equation} \label{eq:EF=FGequiv}
\Big\langle\int_\Sigma F^*\varphi,\int_\Sigma  G^*\tilde\varphi\Big\rangle_{\C^N}=\Big\langle\int_\Sigma G^*\varphi,\int_\Sigma  F^*\tilde\varphi\Big\rangle_{\C^N}\quad (\forall\varphi,\,\tilde\varphi\in L^2(\Sigma;\C^N))
\end{equation} 
that is, in turn, equivalent to the formal symmetry of \eqref{eq:formal_FG}. To see the latter, we note that
\begin{multline*}
\langle(|F\delta_\Sigma\rangle\langle G\delta_\Sigma|\psi),\tilde\psi\rangle\equiv \Big(\overline{F\int_\Sigma (G^*\psi)}\delta_\Sigma,\tilde\psi\Big)=\int_\Sigma\Big\langle F\int_\Sigma (G^*\psi),\tilde\psi\Big\rangle_{\C^N}\\
=\Big\langle\int_\Sigma G^*\psi,\int_\Sigma  F^*\tilde\psi\Big\rangle_{\C^N}
\end{multline*}
and, on the other hand,
\begin{multline*}
\langle\psi,(|F\delta_\Sigma\rangle\langle G\delta_\Sigma|\tilde\psi)\rangle\equiv\overline{\langle(|F\delta_\Sigma\rangle\langle G\delta_\Sigma|\tilde\psi),\psi\rangle}=\overline{\Big\langle\int_\Sigma G^*\tilde\psi,\int_\Sigma  F^*\psi\Big\rangle}_{\C^N}\\
=\Big\langle\int_\Sigma F^*\psi,\int_\Sigma  G^*\tilde\psi\Big\rangle_{\C^N}.
\end{multline*}
Above, we may consider, e.g., $\psi,\,\tilde\psi\in H^1(\R^n;\C^N)$. In that case, the values of $\psi,\,\tilde\psi$ in the integrals over $\Sigma$ should be understood in the sense of traces.  The claim then follows from the facts that the trace mapping maps $H^1(\R^n;\C^N)$ onto $H^{\frac{1}{2}}(\Sigma;\C^N)$ surjectively and that the latter space is dense in $L^2(\Sigma;\C^N)$.
\end{remark}


\begin{example}\label{ex: regular F}
If the columns of $F$ are linearly independent in $L^2(\Sigma;\C^N)$ then the linear mapping $\langle F|_\Sigma:\, L^2(\Sigma;\C^N)\to\C^N$ defined by $\langle F|_\Sigma\varphi:=\int_\Sigma (F^*\varphi)$ is surjective. Therefore,  \eqref{eq:FG=GF}  holds true if and only if  there exists a constant hermitian $N\times N$ matrix $L$ such that $G=F L$. We then have $|F\rangle_\Sigma\langle G|_\Sigma=|F\rangle_\Sigma\langle FL|_\Sigma=:|F\rangle_\Sigma L \langle F|_\Sigma$.
\end{example}

Now, let us inspect spectral properties of $D_{F,G}$. According to the point (i) of Theorem \ref{Theo: Boundary triple}, $z\in\sigma_{\text{p}}(D_{F,G})\setminus\sigma(D_0)$ if and only if 
\begin{equation} \label{eq:ev_cond}
(I+|F\rangle_\Sigma\langle G|_\Sigma M(z))\psi=0
\end{equation}
has a non-zero solution $\psi\in L^2(\Sigma;\C^N)$. Let $\{f_1,f_2,\ldots ,f_N\}$ be columns of $F$, 
\begin{equation*}
\mathcal{F}:=\mathrm{span}\{f_1,f_2,\ldots ,f_N\},
\end{equation*}
and $\{\tilde f_1,\tilde f_2,\ldots,\tilde f_{\tilde N}\}$ a basis of $\mathcal{F}$. Consequently, there exist unique constants $C_{kl}\in\C$ such that
\begin{equation} \label{eq:basis_trafo}
f_k=\sum_{l=1}^{\tilde N} C_{kl}\tilde f_l\quad (k=1,2,\ldots,N).
\end{equation}

Note that $\psi\in\mathcal{F}$ if and only if $(I+|F\rangle_\Sigma\langle G|_\Sigma M(z))\psi\in\mathcal{F}$. In particular, \eqref{eq:ev_cond} yields that 
\begin{equation*}
\psi=\sum_{l=1}^{\tilde N}a_l \tilde f_l
\end{equation*}
for some $a_l\in\C$. Substituting this decomposition back to \eqref{eq:ev_cond} and using \eqref{eq:basis_trafo},  we get
\begin{equation} \label{eq:ef_eq}
\sum_{j=1}^{\tilde N}\sum_{k=1}^N \Big(\int_\Sigma G^* M(z)\tilde f_j\Big)_k C_{kl}\, a_j=-a_l \quad (l=1,2,\ldots, \tilde N),
\end{equation}
where the lower index $k$ denotes the $k$th component of the column vector in the round bracket. Introducing 
\begin{equation} \label{eq:tildeF_C}
\tilde F:=\begin{pmatrix} \tilde f_1 & \tilde f_2&\ldots& \tilde f_{\tilde N}\end{pmatrix}\in L^2(\Sigma;\C^{N\times\tilde N}),\quad C:=(C_{kl})_{k,l=1}^{N,\tilde N}\in\C^{N\times\tilde N},
\end{equation}
we see that the existence of a non-trivial solution $(a_1,a_2,\ldots, a_{\tilde N})^T$ to \eqref{eq:ef_eq} is equivalent to the condition
\begin{equation} \label{eq:BSalt}
\det\Big(I_{\tilde N}+C^T\int_\Sigma(G^*M(z)\tilde F)\Big)=0.
\end{equation}
These considerations prove partially the following theorem.

\begin{theorem}\label{theo spectral cond.}
Let $F,\,G\in L^2(\Sigma;\C^{N\times N})$ be such that \eqref{eq:FG=GF} holds true. Then
$\sigma_{\mathrm{ess}}(D_{F,G})=(-\infty,-|m|]\cup[|m|,+\infty)$
and the number of discrete eigenvalues of $D_{F,G}$ counting multiplicities is at most equal to the number of the linearly independent (in $L^2(\Sigma;\C^N)$) columns of $F$. Furthermore, for $z\in(-|m|,|m|)$,
it holds
\begin{equation} \label{eq:BS}
z\in\sigma_{\mathrm{p}}(D_{F,G}) \quad \text{if and only if} \quad -1\in\sigma\Big(C^T\int_\Sigma(G^*M(z)\tilde F)\Big),
\end{equation}  
where $\tilde F$ and $C$ are given in \eqref{eq:tildeF_C}.
\end{theorem}

\begin{proof}
Firstly, the condition \eqref{eq:BS} is equivalent to \eqref{eq:BSalt} derived above. Next,  in the proof of Theorem \ref{Theo:sa} it was shown that for $z\in\C\setminus\R$ (in fact, for all $z\in\rho(D_0)\setminus\sigma_{\mathrm{p}}(D_{F,G})$), $\Ran(D_{F,G}-z)=L^2(\R^n;\C^N)$. Therefore, applying \eqref{eq:Krein} with $T_0=D_0,\, T_B=D_{F,G},\,$ and $B=|F\rangle_\Sigma\langle G|_\Sigma$ we see that the difference $(D_{F,G}-z)^{-1}-(D_0-z)^{-1}$ is a finite rank operator, because $B$ projects on the $\tilde N$-dimensional space $\mathcal{F}$ spanned by the columns of $F$, $(I+BM(z))$ maps $\mathcal{F}$ onto $\mathcal{F}$ bijectively, and $\gamma(z)$ is bounded from $L^2(\Sigma;\C^N)$ to $L^2(\R^n;\C^N)$. More concretely, the rank of the difference cannot be larger than $\tilde N$. The claim about the essential spectrum then follows from the Weyl criterion and the fact that $\sigma_{\mathrm{ess}}(D_0)=(-\infty,-|m|]\cup[|m|,+\infty)$. The bound on the number of discrete eigenvalues is a consequence of \cite[Chpt. 9.3, Theo. 3]{BiSo}.
\end{proof}

\begin{remark}
It can be easily seen that for $F$ with linearly independent columns we may choose $\tilde{F}=F$ and $C = I_N$. Then, in view of Example \ref{ex: regular F}, the matrix $G$ can be written as $G=FL$, where $L$ is a constant hermitian matrix. Consequently, the spectral condition \eqref{eq:BS} reduces to 
$$z\in\sigma_{\mathrm{p}}(D_{F,G})  \quad \text{if and only if} \quad -1\in\sigma\Big(L\int_\Sigma(F^*M(z) F)\Big).$$
\end{remark}

\section{Non-local approximations} \label{sec:approx}
In this section, we will show that the operator $D_{F,G}$ may be understood as a limit of the free operator $D_0$ with a scaled finite-rank perturbation. Since  we will use the results of Subsection \ref{sec:tubular} to construct the perturbation, let us assume  from now on that $\Omega$ has $C^2$-smooth boundary $\Sigma$. First, note that the tubular $\varepsilon$-neighbourhood of $\Sigma$ introduced in \eqref{eq:epsNeigh} obeys
\begin{equation} \label{eq:epsLayer}
\Sigma_\varepsilon =
  \{x_\Sigma + t \nu(x_\Sigma) \mid x_\Sigma \in \Sigma, \, t
  \in (-\varepsilon,\varepsilon) \}.
\end{equation}  
Here, $t=\varepsilon u$ measures the distance  of a point $x\in\Sigma_\varepsilon$ from $\Sigma$. Since, for all $\varepsilon$ small enough, the representation of $x\in\Sigma_\varepsilon$ given by the right-hand side of \eqref{eq:epsLayer} is unique, we will identify $x$ with the pair $(x_\Sigma,t)$.
Next, let $v \in L^{\infty}(\R;\R)$ be such that $\supp v \subset (-1,1)$ and $\int_{-1}^{1} v(t)\, \dd t = 1$. For $\varepsilon>0$, we put $v_\varepsilon(t):=\varepsilon^{-1}v(\varepsilon^{-1} t)$ and
\begin{equation*}
   F_{\varepsilon}(x) :=
    \begin{cases}
      F(x_\Sigma)v_\varepsilon(t)
      \quad &\text{ for  }x\equiv(x_\Sigma,t)\in\Sigma_\varepsilon
      \\
      0 \quad &\text{ away from } \Sigma_{\varepsilon}.
    \end{cases}
\end{equation*}
A matrix-valued function $G_\varepsilon$ is introduced similarly. It is straightforward to check that, in the sense of distributions, $\lim_{\varepsilon\to 0+}F_\varepsilon=F\delta_\Sigma$. Therefore, the natural candidate for approximating potential is the following finite-rank operator in $L^2(\R^n;\C^N)$,
\begin{equation*}
|F_\varepsilon\rangle\langle G_\varepsilon|\psi:=F_\varepsilon\int_{\R^n} (G_\varepsilon^*\psi)=F_\varepsilon\int_{\Sigma_\varepsilon} (G_\varepsilon^*\psi).
\end{equation*}

Our aim will be to show that
$$D_{F,G}^{\varepsilon}:=D_0+|F_\varepsilon\rangle\langle G_\varepsilon|$$
converges in the norm resolvent sense to $D_{F,G}$ as $\varepsilon\to 0+$. Note that if \eqref{eq:FG=GF} is satisfied, then $|F_\varepsilon\rangle\langle G_\varepsilon|$ is hermitian, and hence $D_{F,G}^{\varepsilon}$ is self-adjoint on $\Dom(D_{F,G}^\varepsilon)=\Dom(D_0)$. For the resolvent of $D_{F,G}^\varepsilon$, we have
\begin{equation} \label{eq:ResId}
(D_{F,G}^{\varepsilon}-z)^{-1}=R_z(I+|F_\varepsilon\rangle\langle G_\varepsilon| R_z)^{-1},
\end{equation}
where $R_z$ should be now understood as a shorthand notation for $(D_0-z)^{-1}$. Of course, \eqref{eq:ResId} is only valid for $z\notin\sigma(D_0)$ such that the inverse on the right-hand side exists. Writing
$$(I+|F_\varepsilon\rangle\langle G_\varepsilon| R_z)^{-1}=(I+|F_\varepsilon\rangle\langle G_\varepsilon| R_z)^{-1}((I+|F_\varepsilon\rangle\langle G_\varepsilon| R_z)-|F_\varepsilon\rangle\langle G_\varepsilon| R_z)$$
we get
\begin{equation} \label{eq:ResIdMod}
(D_{F,G}^{\varepsilon}-z)^{-1}=R_z-R_z(I+|F_\varepsilon\rangle\langle G_\varepsilon| R_z)^{-1}|F_\varepsilon\rangle\langle G_\varepsilon| R_z
\end{equation}

For further calculations we will abandon the \emph{bra-ket} notation and introduce the operator $\Pi:\, L^2(\R^n;\C^N)\to \C^N, \, \psi\mapsto\int_{\R^n}\psi$ instead; so, in particular, we have  $|F_\varepsilon\rangle\langle G_\varepsilon|=F_\varepsilon\Pi G_\varepsilon^*$, where the matrix valued functions $F_\varepsilon$ and $G_\varepsilon^*$ are identified with multiplication operators from $\C^N$ to $L^2(\R^n;\C^N)$ and in $L^2(\R^n;\C^N)$, respectively. We will adopt an analogous convention for the matrix valued functions defined along $\Sigma$. Furthermore, we define the operator $\Pi_\Sigma:\, L^2(\Sigma;\C^N)\to \C^N$ by  $\Pi_\Sigma\psi:=\int_{\Sigma}\psi$.

Now, recall \eqref{eq:tildeF_C} and put $\tilde F_\varepsilon(x):=(\tilde f_1(x_\Sigma) v_\varepsilon(t)\,\, \tilde f_2(x_\Sigma) v_\varepsilon(t)\,\, \ldots \,\, \tilde f_{\tilde N}(x_\Sigma) v_\varepsilon(t))$ for $x\equiv(x_\Sigma,t)\in\Sigma_\epsilon$. Away from $\Sigma_\varepsilon$, we extend $\tilde{f}_i v_\varepsilon$ and $\tilde F_\varepsilon$ by zero.
Then we have
\begin{equation} \label{eq:FepsTrafo}
F_\varepsilon=\tilde F_\varepsilon C^T.
\end{equation}
Also define $\mathcal{F}_\varepsilon:=\mathrm{span}\{\tilde f_1 v_\varepsilon,\, \tilde f_2 v_\varepsilon,\, \ldots ,\, \tilde f_{\tilde N} v_\varepsilon\}$ and the mapping $P_\varepsilon: \tilde f_i v_\varepsilon\mapsto \tilde f_i$, which extends by linearity to an isomorphism from $\mathcal{F}_\varepsilon$ onto $\mathcal{F}$. 
In the following, assume that the basis $\{\tilde f_i\}_{i=1}^{\tilde N}$ of $\mathcal{F}$ is orthonormal in $L^2(\Sigma;\C^N)$. With that choice, the mapping 
$$\mathcal{P}:\, \mathcal{F}\to \C^{\tilde N},\quad \psi\mapsto  (\langle \tilde f_1 ,\psi\rangle_\Sigma,\, \langle \tilde f_2,\psi\rangle_\Sigma,\, \ldots,\, \langle \tilde f_{\tilde N},\psi\rangle_\Sigma)^T$$
obeys 
\begin{equation} \label{eq:PId}
\tilde F \mathcal{P}=I_{\mathcal{F}},\quad \mathcal{P}\tilde F=I_{\C^{\tilde N}},\quad \tilde F_\varepsilon \mathcal{P}P_\varepsilon=I_{\mathcal{F}_\varepsilon},\quad \mathcal{P}P_\varepsilon\tilde F_\varepsilon=I_{\C^{\tilde N}}.
\end{equation}

Using the latter two equalities and \eqref{eq:FepsTrafo} we obtain
\begin{multline} \label{eq:ResTrick}
(I_{\mathcal{F}_\varepsilon}+F_\varepsilon\Pi G_\varepsilon^* R_z)^{-1}F_\varepsilon=\tilde{F}_\varepsilon\mathcal{P}P_\varepsilon(\tilde{F}_\varepsilon(I_{\C^{\tilde N}}+ C^T\Pi G_\varepsilon^* R_z\tilde{F}_\varepsilon)\mathcal{P}P_\varepsilon)^{-1}\tilde{F}_\varepsilon C^T\\
=\tilde{F}_\varepsilon(I_{\C^{\tilde N}}+ C^T\Pi G_\varepsilon^* R_z\tilde{F}_\varepsilon)^{-1} C^T
\end{multline}
on $\C^N$. Since $(I+|F_\varepsilon\rangle\langle G_\varepsilon|R_z)$ maps into $\mathcal{F}_\varepsilon$ exactly those vectors that belong to $\mathcal{F}_\varepsilon$, we may combine \eqref{eq:ResIdMod} and \eqref{eq:ResTrick} to get
\begin{equation} \label{eq:resFGApprox}
(D_{F,G}^{\varepsilon}-z)^{-1}=R_z-R_z\tilde{F}_\varepsilon(I_{\C^{\tilde N}}+ C^T\Pi G_\varepsilon^* R_z\tilde{F}_\varepsilon)^{-1} C^T\Pi G_\varepsilon^* R_z.
\end{equation}
The formula holds for all $z\notin\sigma(D_0)$ such that the $\tilde N\times\tilde N$-matrix
$I_{\C^{\tilde N}}+ C^T\Pi G_\varepsilon^* R_z\tilde{F}_\varepsilon$
is invertible. Note that this implies that $z\in(-|m|,|m|)$ belongs to the discrete spectrum of $D_{F,G}^\varepsilon$ if and only if 
\begin{equation} \label{eq:spectralCondApprox}
\det(I_{\C^{\tilde N}}+ C^T\Pi G_\varepsilon^* R_z\tilde{F}_\varepsilon)=0.
\end{equation}

To find the resolvent of $D_{F,G}$, we use \eqref{eq:Krein} with $B=F\Pi_\Sigma G^*$ which yields
\begin{equation*}
(D_{F,G}-z)^{-1}=R_z-\gamma(z)(I+F\Pi_\Sigma G^* M(z))^{-1}F\Pi_\Sigma G^*\gamma(\bar z)^*.
\end{equation*}
With the help of the first two equalities in \eqref{eq:PId} together with $F=\tilde F C^T$ we may use similar manipulations as in \eqref{eq:ResTrick} to rewrite this as follows
\begin{equation} \label{eq:resFG}
(D_{F,G}-z)^{-1}=R_z-\gamma(z)\tilde{F}(I_{\C^{\tilde N}}+C^T\Pi_\Sigma G^* M(z)\tilde F)^{-1}C^T\Pi_\Sigma G^*\gamma(\bar z)^*.
\end{equation}
We are now prepared to state and prove the main result of this section.

\begin{theorem} \label{theo:approx}
Let $\Omega$ have $C^2$-smooth boundary and $F,G\in L^2(\Sigma;\C^{N\times N})$ be such that \eqref{eq:FG=GF} holds true. Then for every $z\notin\sigma(D_{F,G})$ there exists $\varepsilon_z>0$ such that for all $\varepsilon\in(0,\varepsilon_z)$, $z\notin\sigma(D_{F,G}^\varepsilon)$ and
\begin{equation*}
\lim_{\varepsilon\to 0+}\|(D_{F,G}-z)^{-1}-(D_{F,G}^\varepsilon-z)^{-1}\|=0.
\end{equation*}
\end{theorem}

\begin{proof}
Using \eqref{eq:resFGApprox}, \eqref{eq:resFG}, and the joint continuity of the operator composition we get
\begin{equation*}
\|(D_{F,G}-z)^{-1}-(D_{F,G}^\varepsilon-z)^{-1}\|\leq const.\,( C_1(\varepsilon)+C_2(\varepsilon)+C_3(\varepsilon)),
\end{equation*}
whenever the functions
\begin{align*}
&C_1(\varepsilon):=\|\gamma(z)\tilde{F}-R_z\tilde{F}_\varepsilon\|_{\C^{\tilde N}\to L^2(\R^n;\C^N)},\\
&C_2(\varepsilon):=\|(I_{\C^{\tilde N}}+C^T\Pi_\Sigma G^* M(z)\tilde F)^{-1}-(I_{\C^{\tilde N}}+ C^T\Pi G_\varepsilon^* R_z\tilde{F}_\varepsilon)^{-1}\|_{\C^{\tilde N}\to\C^{\tilde N}},\\
&C_3(\varepsilon):=\|C^T(\Pi_\Sigma G^*\gamma(\bar z)^*-\Pi G_\varepsilon^* R_z)\|_{L^2(\R^n;\C^N)\to\C^{\tilde N}}
\end{align*}
are bounded on a right neighbourhood of $\varepsilon=0$. We are going to show that, for $i\in\{1,2,3\}$, $\lim_{\varepsilon\to 0+}C_i(\varepsilon)=0$.
Note that it is sufficient for $C_2(\varepsilon)$ and $C_3(\varepsilon)$ to converge to zero as $\varepsilon\to 0+$ that
\begin{equation*}
\tilde C_2(\varepsilon):=\|\Pi_\Sigma G^* M(z)\tilde F-\Pi G_\varepsilon^* R_z\tilde{F}_\varepsilon\|_{\C^{\tilde N}\to\C^N}
\end{equation*}
and
\begin{equation*}
\tilde C_3(\varepsilon):=\|\Pi_\Sigma G^*\gamma(\bar z)^*-\Pi G_\varepsilon^* R_z\|_{L^2(\R^n;\C^N)\to\C^N},
\end{equation*}
tend to zero as $\varepsilon\to 0+$, respectively. Moreover, if $\lim_{\varepsilon\to 0+}\tilde C_2(\varepsilon)=0$ then the spectral conditions \eqref{eq:spectralCondApprox} and \eqref{eq:BS} together with  the continuity of the determinant with respect to any matrix norm yield the first statement of the theorem.

First, we will investigate the term $C_1(\varepsilon)$. Since $\int_{-1}^1 v=1$, we have
\begin{equation*}
(\gamma(z)\tilde{F})(x)=\int_{-1}^{1}\int_\Sigma R_z(x-y_\Sigma)\tilde{F}(y_\Sigma)v(u)\dd\sigma(y_\Sigma)\dd u \quad (\forall x\in\R^n\setminus\Sigma).
\end{equation*}
Using $\supp(\tilde{F}_\varepsilon)\subset\overline{\Sigma_\varepsilon}$, the parallel coordinates $(x_\Sigma,u)$ introduced in Section \ref{sec:tubular}, and $t=\varepsilon u$, we get
\begin{equation*}
(R_z\tilde{F}_\varepsilon)(x)=\int_{-1}^1\int_\Sigma R_z(x-y_\Sigma-\varepsilon u \nu(y_\Sigma))\tilde{F}(y_\Sigma)v(u) w_\varepsilon(y_\Sigma,u)\dd\sigma(y_\Sigma)\dd u
\end{equation*}
for a.e. $x\in\R^n$. Given $a\in\C^{\tilde N}$, we may estimate as follows
\begin{multline} \label{eq:Aest}
\|(\gamma(z)\tilde{F}-R_z\tilde{F}_\varepsilon)a\|_{L^2(\R^n;\C^N)}=\|(A_0-A_\varepsilon)(\sum_{i=1}^{\tilde N}a_i\tilde{f}_i v)\|_{L^2(\R^n;\C^N)}\\
\leq \|(A_0-A_\varepsilon)\|_{\mathscr{H}\to L^2(\R^n;\C^N)}\tilde{N}\max_{i}\|\tilde{f}_i v\|_\mathscr{H}\max_{j}|a_j|,
\end{multline}
where $\mathscr{H}:=L^2(\Sigma\times(-1,1),\dd\sigma\dd u;\C^N)$ and, for $\psi\in\mathscr{H}$, the operators $A_0$ and $A_\varepsilon$ are defined by
\begin{align*}
&(A_0\psi)(x):=\int_{-1}^{1}\int_\Sigma R_z(x-y_\Sigma)\psi(y_\Sigma,u)\dd\sigma(y_\Sigma)\dd u, \\
&(A_\varepsilon\psi)(x):=\int_{-1}^1\int_\Sigma R_z(x-y_\Sigma-\varepsilon u \nu(y_\Sigma))\psi(y_\Sigma,u) w_\varepsilon(y_\Sigma,u)\dd\sigma(y_\Sigma)\dd u.
\end{align*}  
It was proved in \cite[Prop. 3.8]{BeHoSt_23} that $\lim_{\varepsilon\to 0+}\|(A_0-A_\varepsilon)\|_{\mathscr{H}\to L^2(\R^n;\C^N)}=0$. In view of \eqref{eq:Aest}, we conclude that $\lim_{\varepsilon\to 0+}C_1(\varepsilon)=0$.

Next, we will look at the term $\tilde{C}_3(\varepsilon)$. Since, for all $\psi\in L^2(\R^n;\C^N)$ and a.e. $x_\Sigma\in\Sigma$, 
\begin{equation*}
(\gamma(\bar{z})^*\psi)(x_\Sigma)=\int_{\R^n}R_z(x_\Sigma-y)\psi(y)\dd y,
\end{equation*}
and $\int_{-1}^{1}v=1$, we deduce that
\begin{equation*}
\Pi_\Sigma G^*\gamma(\bar z)^*\psi=\int_{-1}^1\int_{\Sigma}G^*(x_\Sigma)v(u)\int_{\R^n}R_z(x_\Sigma-y)\psi(y)\dd y\dd\sigma(x_\Sigma)\dd u.
\end{equation*}
Furthermore, using a similar reasoning as in the previous paragraph, we obtain
\begin{multline*}
\Pi G_\varepsilon^* R_z\psi\\
=\int_{-1}^1\int_{\Sigma}G^*(x_\Sigma)v(u)\int_{\R^n}R_z(x_\Sigma+\varepsilon u \nu(x_\Sigma)-y)\psi(y)\dd y \,w_\varepsilon(x_\Sigma,u)\dd\sigma(x_\Sigma)\dd u.
\end{multline*}
If we introduce bounded operators $C_0,C_\varepsilon:\, L^2(\R^n;\C^N)\to\mathscr{H}$ as follows
\begin{align*}
& (C_0\psi)(x_\Sigma,u):=\int_{\R^n}R_z(x_\Sigma-y)\psi(y)\dd y,\\
& (C_\varepsilon\psi)(x_\Sigma,u):=\int_{\R^n}R_z(x_\Sigma+\varepsilon u \nu(x_\Sigma)-y)\psi(y)\dd y,
\end{align*}
then for the $i$th component of  $(\Pi_\Sigma G^*\gamma(\bar z)^*-\Pi G_\varepsilon^* R_z)\psi$ we get
\begin{multline*}
\mkern-18mu \left((\Pi_\Sigma G^*\gamma(\bar z)^*-\Pi G_\varepsilon^* R_z)\psi\right)_i=\int_{-1}^1\int_\Sigma \bar{g_i}^T(x_\Sigma)v(u)(C_0\psi-w_\varepsilon C_\varepsilon\psi)(x_\Sigma,u)\dd \sigma(x_\Sigma)\dd t\\
=\int_{-1}^1\int_\Sigma \bar{g_i}^T(x_\Sigma)v(u)w_\varepsilon(x_\Sigma,u)(C_0\psi-C_\varepsilon\psi)(x_\Sigma,u)\dd \sigma(x_\Sigma)\dd t\\
+\int_{-1}^1\int_\Sigma (1-w_\varepsilon(x_\Sigma,u))\bar{g_i}^T(x_\Sigma)v(u)C_0\psi(x_\Sigma,u)\dd\sigma( x_\Sigma)\dd t,
\end{multline*}
where $g_i$ stands for the $i$th column of the matrix $G$. Applying the Cauchy-Schwarz inequality together with \eqref{eq:wAsy}, we deduce that
\begin{multline*}
|\left((\Pi_\Sigma G^*\gamma(\bar z)^*-\Pi G_\varepsilon^* R_z)\psi\right)_i|\\
\leq (1+\mathcal{O}(\varepsilon))\|g_i v\|_{\mathscr{H}}\|C_0-C_\varepsilon\|_{ L^2(\R^n;\C^N)\to\mathscr{H}}\|\psi\|_{L^2(\R^n;\C^N)}\\
+\mathcal{O}(\varepsilon)\|g_i v\|_{\mathscr{H}}\|C_0\|_{ L^2(\R^n;\C^N)\to\mathscr{H}}\|\psi\|_{L^2(\R^n;\N)}.
\end{multline*}
By \cite[Prop. 3.7]{BeHoSt_23}, $\lim_{\varepsilon\to 0+}\|C_0-C_\varepsilon\|_{ L^2(\R^n;\C^N)\to\mathscr{H}}=0$. Consequently, $\tilde{C}_3(\varepsilon)$ also converges to zero as $\varepsilon\to 0+$.

Finally, we will be concerned with the term $\tilde{C}_2(\varepsilon)$.
For the $(i,j)$th element of the matrix $\Pi G_\varepsilon^* R_z\tilde{F}_\varepsilon$ we find that
\begin{equation*}
(\Pi G_\varepsilon^* R_z\tilde{F}_\varepsilon)_{ij}=\int_{-1}^1\int_\Sigma \bar{g_i}^T(x_\Sigma)v(u) (B_\varepsilon(\tilde{f}_j v))(x_\Sigma,u)w_\varepsilon(x_\Sigma,u)\dd\sigma(x_\Sigma)\dd u,
\end{equation*}
where
\begin{equation*}
 (B_\varepsilon\psi)(x_\Sigma,u):=\int_{-1}^1\int_\Sigma R_z(x_\Sigma+\varepsilon u\nu(x_\Sigma)-y_\Sigma-\varepsilon s\nu(y_\Sigma))\psi(y_\Sigma,s)w_\varepsilon(y_\Sigma,s)\dd \sigma(y_\Sigma)\dd s
\end{equation*}
is a bounded operator in $\mathscr{H}$. It follows from \cite[Prop. 3.10]{BeHoSt_23} that $B_\varepsilon$ converges to  $B_0$ in the space of bounded operators from $\mathscr{H}_{1/2}:=L^2((-1,1);H^{1/2}(\Sigma;\C^N))$ to $\mathscr{H}$, where $B_0$ defined by
\begin{equation*}
(B_0\psi)(\cdot, u):=\frac{i}{2}(\alpha\cdot \nu)\int_{-1}^1\sgn(u-s)\psi(\cdot,s)\dd s+M(z)\int_{-1}^1\psi(\cdot,s)\dd s
\end{equation*}
is a bounded operator in $\mathscr{H}$. By density, for every $\psi\in\mathscr{H}$ we find a sequence $(\psi_n)\subset\mathscr{H}_{1/2}$ that converges to $\psi$ in $\mathscr{H}$. Therefore, we have
\begin{multline*}
\|(B_\varepsilon-B_0)\psi\|_{\mathscr{H}}\leq \|(B_\varepsilon-B_0)\psi_n\|_{\mathscr{H}}+\|(B_\varepsilon-B_0)(\psi-\psi_n)\|_{\mathscr{H}}\\
\leq \|B_\varepsilon-B_0\|_{\mathscr{H}_{1/2}\to\mathscr{H}}\|\psi_n\|_{\mathscr{H}_{1/2}}+(\|B_\varepsilon\|_{\mathscr{H}\to\mathscr{H}}+\|B_0\|_{\mathscr{H}\to\mathscr{H}})\|\psi-\psi_n\|_{\mathscr{H}}.
\end{multline*}
Since again by \cite[Prop. 3.10]{BeHoSt_23} the operators $B_\varepsilon$ are uniformly bounded in $\mathscr{H}$, we infer that $B_\varepsilon$ converges to $B_0$ strongly in $\mathscr{H}$ as $\varepsilon\to 0+$. Furthermore, due to \eqref{eq:wAsy}, we see that $\lim_{\varepsilon\to 0+}(g_i v w_\varepsilon)=g_i v$ in $\mathscr{H}$. Using these two results together with the  joint continuity of the dot product and the fact that $\int_{-1}^1 v=1$, we arrive at
\begin{multline*}
\lim_{\epsilon\to 0+}(\Pi G_\varepsilon^* R_z\tilde{F}_\varepsilon)_{ij}=\int_{-1}^1\int_\Sigma \bar{g_i}^T(x_\Sigma)v(u) (B_0(\tilde{f}_j v))(x_\Sigma,u)\dd\sigma(x_\Sigma)\dd u\\
=\frac{i}{2}\int_{-1}^1\int_{-1}^1v(u)\sgn(u-s)v(s)\dd u\dd s \int_\Sigma \bar{g_i}^T(x_\Sigma)(\alpha\cdot \nu(x_\Sigma))\tilde{f}_j(x_\Sigma)\dd\sigma(x_\Sigma)\\
+\int_\Sigma \bar{g_i}^T(x_\Sigma) (M(z)\tilde{f}_j)(x_\Sigma)\dd\sigma(x_\Sigma).
\end{multline*}
The first term on the right-hand side is clearly zero, whereas the second one is just the integral representation of the $(i,j)$th element of $\Pi_\Sigma G^*M(z)\tilde{F}$. Hence, we conclude that $\lim_{\varepsilon\to 0+}\tilde{C}_2(\varepsilon)=0$.
\end{proof}

\begin{remark}
Essentially the same operators as $A_\varepsilon,\, B_\varepsilon,$ and $C_\varepsilon$ from the proof of Theorem \ref{theo:approx} were originally studied in \cite{Mas Piz} for $n=3$. Although the convergence results obtained there are weaker than the results from the recent preprint \cite{BeHoSt_23}, they are still strong enough to support our proof. In fact, they may be generalized to the dimension $n=2$ in a rather straightforward way and then used in our proof, too. 
\end{remark}

\section*{Acknowledgment}
L. Heriban acknowledges  the support by the EXPRO grant No. 20-17749X of the Czech Science Foundation (GA\v{C}R). M.~Tu\v{s}ek was partially supported by the grant No.~21-07129S of the Czech Science Foundation (GA\v{C}R).

The authors thank Markus Holzmann for fruitful discussions.

\end{document}